\documentstyle[12pt]{article}
\newlength{\bredde}
\def\slash#1{\settowidth{\bredde}{$#1$}\ifmmode\,\raisebox{.15ex}{/}
\hspace*{-\bredde} #1\else$\,\raisebox{.15ex}{/}\hspace*{-\bredde} #1$\fi}
\newcommand{\beq}{\begin{equation}}
\newcommand{\eeq}{\end{equation}}

\newcommand{\lG}{\raise.3ex\hbox{$\stackrel{\leftarrow}{G}$}}
\newcommand{\lU}{\raise.3ex\hbox{$\stackrel{\leftarrow}{U}$}}
\newcommand{\lP}{\raise.3ex\hbox{$\stackrel{\leftarrow}{{\cal P}}$}}
\newcommand{\leta}{\raise.3ex\hbox{$\stackrel{\leftarrow}{\eta}$}}
\newcommand{\lOmega}{\raise.3ex\hbox{$\stackrel{\leftarrow}{\Omega}$}}
\newcommand{\ldr}{\raise.3ex\hbox{$\stackrel{\leftarrow}{\delta^r}$}}
\def\beqn{\begin{eqnarray}}
\def\eeqn{\end{eqnarray}}
\def\sepand{\rule{14cm}{0pt}\and}
\def\gtwid{\raise.3ex\hbox{$>$\kern-.75em\lower1ex\hbox{$\sim$}}}
\def\ltwid{\raise.3ex\hbox{$<$\kern-.75em\lower1ex\hbox{$\sim$}}}

\begin{document}
\topmargin -1.4cm
\oddsidemargin -0.8cm
\evensidemargin -0.8cm
\title{\Large{Constraints on Beta Functions from 
Duality}\footnote{This work is supported in part by funds
provided by the U.S. Department of Energy (D.O.E.) under
cooperative research agreement \#DF-FC02-94ER40818,
and by NSF Grant PHY-92-06867.}}

\vspace{0.9cm}

\author{
{\sc Poul H. Damgaard}\\
The Niels Bohr Institute\\ Blegdamsvej 17\\ DK-2100 Copenhagen\\
Denmark\\
\sepand
{\sc Peter E. Haagensen}\\
Center for Theoretical Physics\\
Laboratory for Nuclear Science\\
and Department of Physics\\
Massachusetts Institute of Technology\\
Cambridge, MA 02139}
 
\maketitle
\vfill
\begin{abstract} 
We analyze the way in which duality constrains the exact
beta function and correlation length in single-coupling spin
systems. A consistency condition we propose shows very concisely
the relation
between self-dual points and phase transitions, and implies
that the correlation length must be duality invariant.
These ideas are then tested on the 2-d Ising model, and used
towards finding the exact beta function of the $q$-state 
Potts model. Finally, a generic procedure 
is given for identifying
a duality symmetry in other single-coupling models with
a continuous phase transition.

\end{abstract}
\vfill
\begin{flushleft}
NBI-HE-96-50 \\
MIT-CTP \#2569\\
cond-mat/9609242
\end{flushleft}
\thispagestyle{empty}
\newpage

\setcounter{page}{1}

Self-duality plays an important r\^{o}le in both statistical mechanics and
quantum field theory \cite{reviews}. In its simplest manifestation, it 
relates the partition function of a theory with one coupling constant $K$ to 
the partition function of the same theory evaluated at a different
coupling $\tilde{K}$. The most well-known example is that of
Kramers-Wannier duality in the 2-dimensional Ising model \cite{KW}, 
which has various generalizations to different spin models.

Very recently, self-duality has been a topic of intense studies in the
context of field theory and string theory. A particularly important case 
is that of
2-dimensional $\sigma$-models, where a so-called T-duality can be attached 
to both abelian \cite{B} and non-abelian \cite{OQ} isometries of the 
target manifold. In this connection, the focus has
most often been on conformally invariant backgrounds, because these
are the backgrounds of main interest from the point of
view of string theory. But in these $\sigma$-model examples duality
has been thought to be of wider applicability, beyond conformal
points. This prompted one of us to investigate the constraints implied
by duality on renormalization group (RG) flows in these $\sigma$-models
\cite{H}. One of the main lessons learned from this analysis is that
duality forces certain consistency conditions on the $\sigma$-model
beta functions. In Ref. \cite{H}
these conditions were shown to independently determine the
one-loop beta functions (as well as the expected 
dilaton shift) in a bosonic $\sigma$-model with an abelian
isometry.

In this Brief Report we wish to point out that such consistency
conditions are of relevance in the context of statistical mechanics 
as well.
For instance, we find that they provide an extremely 
transparent explanation
for the close relationship between self-dual points $K^*$
and phase transitions. They also predict exact duality invariance
of the correlation length. 
In fact, starting from a fairly general setting, we find that 
duality yields a number of results which elucidate 
the RG and phase structure of different models, and this 
may in turn lead to definite predictions on observables in 
these models. 
In our effort to test these ideas, we are aided by the fact
that a full solution of the 2-dimensional Ising model on a
square lattice is known \cite{O}, while we are on the other hand 
limited by the scarcity of exact results on most other 
models presenting duality.

Consider first a theory, for definiteness a spin model, with just one
coupling constant $K$. If this theory is self-dual, there exists a
relation between this coupling and its dual $\tilde{K}$,
\beq
\tilde{K} ~=~ t(K) ~,\label{dualK}
\eeq
such that the partition functions at $K$ and $\tilde{K}$ are
related in some simple way, which shall not presently be of 
concern to us. The only essential requirement is that the interaction
parts of the Hamiltonian and its dual are of precisely the same
form, with $K$ replaced by $\tilde{K}$.
We assume that the duality map is continuous and $1\! :\! 1$, and that it 
gives back the original coupling when acting twice:
\beq
t(t(K)) ~=~ K ~.
\label{tt}\eeq
This implies $t'(K) < 0$, and we 
assume $K, \tilde{K} > 0$. At the self-dual point
$K^*$ we have $t(K^*) = K^*$.

In quantum field theory, such duality relations can be given meaning
only at a given scale of the ultraviolet cut-off. In the present spin
model context, the ultraviolet cut-off is automatically included in
the definition of the model ($e.g$, square lattice, triangular lattice, 
etc.). We shall now explore some simple implications these duality
relations have on the RG flow. This flow is generated by an infinitesimal
change in the ultraviolet cut-off, here as a change in the lattice
spacing $a \to a + \delta a$. In the conventional manner, this leads
us to define a beta function by
\beq
\beta(K) ~=~ -a \frac{\partial K}{\partial a} ~,\label{betafunction}
\eeq
which stipulates how the coupling constant $K$ is renormalized under 
a change of scale.

The consistency condition on $\beta(K)$ is arrived at 
by first applying the operator $-a\partial/\partial a$ to both 
sides of the duality relation, Eq.~(\ref{dualK}):
\beq
-a\frac{\partial}{\partial a} \tilde{K} ~=~ -a\frac{\partial K}{\partial a}
\frac{\partial \tilde{K}}{\partial K} ~=~ \beta(K)t'(K) ~.
\label{chain}\eeq 
We now impose that the left hand side of this equation
be identified with the beta function itself, $\beta(\cdot )$, evaluated
at $\tilde{K}$: 
\beq
\beta(\tilde{K})= \beta(K)t'(K)\ .
\label{consist}\eeq
This is our consistency condition for the beta function.
Insofar as $\tilde{K}$ is some function of $K$, it would seem 
that its RG flow should simply be dictated by that of $K$.
What we are proposing, however, is a stronger condition:
the requirement in Eq.~(\ref{consist})
that the flows of $K$ and $\tilde{K}$
be thus ``tied together", as it were, imposes stringent 
restrictions on what this flow can actually be. A simple
example below will suffice to show this clearly.

An immediate result that follows is a general relationship
between self-dual points and phase transitions. At $K=K^*$,
Eq. (\ref{tt}) implies that $t'(K^*)=-1$, and thus there
are only two ways of satisfying Eq.~(\ref{consist}) at
$K=K^*$: either
\begin{itemize}
\item the beta function is continuous at $K = K^*$, and $\beta(K^*)
 = 0$ \, , or
\item the beta function is discontinuous at $K = K^*$, with 
$\beta(K\!\uparrow\! K^*) = -\beta(K\!\downarrow\! K^*)$ ~. 
\end{itemize}
The former case applies to second and higher order phase transitions,
or to high/low temperature sinks where the correlation length
vanishes. 
The latter possibility, a symmetric discontinuity around zero
in the beta function, has perhaps a less obvious interpretation.
We view it as pertaining to first order phase transitions. This
is consistent with the discontinuity in the correlation length
at such transitions, and can also be understood 
in terms of co-existing phases. Namely, we may consider the
RG flows on both sides of the first order phase transition, and
their continuation within one phase only on the other side of
the transition. Such analytic continuations will typically converge
towards a zero close to the phase transition. What duality symmetry
enforces, then, is that the two continuations occur symmetrically
from either side of the phase transition, so that the two
branches end with equal magnitude and opposite signs just at the
phase transition point itself. 
We recall that Kramers and Wannier
originally identified the self-dual point in the Ising model
as a phase transition point by associating it to a point of 
possible non-analyticity in the free energy  \cite{KW}. 
We do so in greater generality
by associating it with a point of onset of scale invariance,
with the advantage that our derivation need make no assumptions
about the {\it lack} of phase transition points elsewhere. We now consider 
what Eq.~({\ref{consist}) entails for the correlation length.

Under the RG rescaling $a \to a' \equiv 
a + \delta a$, the correlation length changes according to
$\xi(K') = \xi(K)(1 + \delta a/a)$. Using Eq.~(\ref{betafunction}),
this gives the well-known relation
\beq
\beta(K) ~=~ - \frac{\xi(K)}{\xi'(K)} ~. \label{betacorr}
\eeq
Together with Eq.~({\ref{consist}), this then implies that the correlation 
length
should be invariant under duality transformations:
\beq
\xi (\tilde{K})=\xi(K)\ .
\label{xi}\eeq
Although we are interested here mostly in systems with finite order
phase transitions, it is clear that Eqs.~(\ref{consist})-(\ref{xi})
can be considered also in the context of systems with more exotic 
behavior, as for instance the 2-d XY model.

Let us now consider the 2-dimensional Ising model on a square lattice
(in the absence of external magnetic field) as a specific example.
In this case the exact beta function can be extracted from Onsager's 
\cite{O} original
solution (see Ref. \cite{Ba}). In a conventional normalization the
Hamiltonian is taken to be $H = K \sum s_i s_j$, where the sum runs over
all nearest-neighbor spins. The duality relation can then be
written \cite{KW}:
\beq
\tanh K ~=~ e^{-2\tilde{K}} ~,\label{Isingdual}
\eeq  
which yields the well-known expression $K^* = (1/2)\ln(1 + \sqrt{2}) =
0.440....$ for the self-dual point. Moreover, from the exact expression
for the infinite-volume correlation length \cite{FB}
\beq
\xi(K) ~=~\frac{1}{|2K+\ln\tanh K|}=\frac{1}{2}\ \frac{1}{|K-t(K)|} ~,
\label{xiIsing}\eeq
one finds, from Eq. (\ref{betacorr}), the exact beta function:
\beq
\beta(K)=
\frac{1}{2}\ \frac{\sinh 2K}{1 + \sinh 2K}\ (2K +\ln\tanh K) 
=\frac{K-t(K)}{1-t'(K)}\ .\label{betaIsing}
\eeq
As written above, it is easy to see that
the beta function does indeed satisfy the consistency condition,
and that the correlation length is (manifestly) duality invariant.
At this point one can see more explicitly how restrictive the consistency 
condition is: using, for instance, Ising duality, Eq.~(\ref{Isingdual}), 
if one were to
{\it arbitrarily} choose some ``flow" for $K$, say $K(a)=1/a$ 
(which would yield $\beta(K)=K$), then Eq.~(\ref{chain}), simply
expressing a chain rule would, naturally, be satisfied, while
the consistency condition, Eq.~(\ref{consist}), would 
certainly not.

When an
explicit lattice construction of the duality map is available through
an associated operator transformation from the lattice to its
dual, duality invariance of the correlation length can often
be checked explicitly from the 2-point correlation function.
This is, for example, the case of the 2-d Ising model. 
However, our general considerations are of course
not limited to these special cases. We find, furthermore,
the present derivation to be more direct and transparent.

While this Ising model solution is completely known, the generalization
to the $q$-state Potts model on a square lattice provides interesting
ground for conjectures, since there duality is also known, but the
exact beta function (or correlation length) is not. In these models,
the duality transformations, $\tilde{K}=t_q(K)$, are most simply
expressed implicitly through \cite{Wu}
\beq
(e^{qK} - 1)(e^{q\tilde{K}} - 1) ~=~ q ~,\label{Pottsdual}
\eeq
which reduces to Eq.~(\ref{Isingdual}) for $q\!=\!2$, as it should. 
In general, for continuous phase transitions with a power-law 
divergence of the correlation length near the critical point $K_c$,
we have 
\beq
\xi(K) ~=~ A|K - K_c|^{-\nu} + \ldots ~,
\eeq
and, through use of Eq.~(\ref{betacorr})
the beta function develops a simple zero,
\beq
\beta(K) ~=~ \frac{1}{\nu}(K - K_c) + \ldots ~.
\eeq
This also recovers the well-known fact that the slope of the beta function
at $K_c$ is determined by the critical exponent $\nu = 1/\beta'(K_c)$.
What we have seen is that in the Ising case (where $\nu=1$), the 
behavior above 
is indeed exact for all couplings
if one substitutes $|K - K_c|\rightarrow |K - \tilde{K}|$. If we
assume that the same also happens in the generalization to $q\not= 2$,
we are led to predict that the exact correlation length and beta function
are
\beq
\xi_q(K)=A_q|K -t_q(K)|^{-\nu_q}\eeq
\beq
\beta_q(K)=\frac{1}{\nu_q}\ \frac{K-t_q(K)}{1-t'_q(K)}\ ,
\eeq
where $\nu_q\!=\! (8-q)/6$ for $q\!=\! 2,3,4$ .
However, one can see that the solution to Eq.~(\ref{consist}), or
Eq.~(\ref{xi}), is not unique given only $t(K)$ and the critical behavior
at $K^*$,
and thus, without further external input, we are not quite ready
yet to make such a strong conjecture. Furthermore, it is also well-known
that for $q\geq 5$, the transition becomes first order, with a finite 
correlation length at the phase transition. With the beta function as related
in Eq.~(\ref{betacorr}) to the correlation length, it follows that it must 
then satisfy the consistency condition with a discontinuity at the 
critical point, which is not the case for $\beta_q(K)$ above. We feel,
however, that the above considerations will be useful in determining
exact results, both for $q\!=\!3,4$ and $q\geq5$, once more information 
becomes available in these models. In particular, in view of these
considerations, numerical or analytical determinations of these
beta functions would be highly interesting.

We turn now to a question which is motivated by the general
character of the Ising model beta function and correlation length:
as given by Eq.~(\ref{xiIsing}), $\xi (K)$ starts off at zero, grows 
analytically
to infinity at $K\!=\!K^*$, comes down to the right of that and asymptotes 
to zero, again analytically. Accordingly, the beta function also starts off
at zero, becomes negative, crosses zero at $K\!=\!K^*$, becoming positive 
and diverging linearly at infinity.  We now ask the question:
given any other model in which the
correlation length or beta function behave
essentially in this way, is it reasonable to expect that it
will also present some form of duality? Or, given $\beta (K)$
or $\xi (K)$, can one find the duality transformation $t(K)$ uniquely?
In fact, one can. We will construct such a $t(K)$ here in two alternative 
ways:
first, by a ``Taylor reconstruction" procedure, and then by a graphical 
procedure. 

The first procedure consists in simply expanding 
Eq.~(\ref{consist}) in a Taylor series around the self-dual point. To
that end, we first note that whenever $\xi (K)$
goes to zero or infinity analytically at a certain finite $K$, the beta 
function has an analytical zero there. This means in particular that
a Taylor series around the self-dual point makes sense. 
Moreover, infinitesimally close to the self-dual point,
$K$ and $\tilde{K}$ are also infinitesimally close.
Thus, if we
perform such an expansion, there will be, order by order in the 
expansion parameter, a relation giving derivatives of $t(K)$
at $K^*$ in terms of derivatives of $\beta (K)$ and lower derivatives
of $t(K)$, also at $K^*$. This procedure is straightforward
and systematic, and we will not present it explicitly here.
The main implication is that we can compute the duality map $t(K)$
order by order in a Taylor expansion around $K^*$. This (unique) duality map
will by construction leave the correlation length $\xi(K)$ invariant,
while the beta function will also satisfy the
consistency condition, Eq.~(\ref{consist}). Of course, we cannot from this
alone argue that $t(K)$ is a genuine duality transformation in the
sense of a mapping between two dual Hamiltonians (and hence all
observables). But the existence of such a map $t(K)$ in this generalized
situation is highly suggestive. It would be quite interesting to
check by Monte Carlo methods whether the resulting map $t(K)$ indeed
represents a full duality transformation on other observables. 
One other point is noteworthy:
the above procedure cannot be turned around to yield
$\beta (K)$ from $t(K)$, because some derivatives of 
$\beta (K)$ are simply free and drop out of the equations.
Of course, this is consistent with the fact that there are many 
beta functions which solve Eq.~(\ref{consist}) given $t(K)$,
so that there cannot be a unique reconstruction procedure
for $\beta (K)$. 

The alternative procedure consists in building $t(K)$ from the 
$\xi (K)$ profile: call $\xi_L (K)$ and $\xi_R (K)$
the profiles of $\xi (K)$ to the left and to the right of the
self-dual point, respectively. If the inverse functions $\xi^{-1}_R$
and $\xi^{-1}_L$ exist, 
the duality transformation 
$t (K)$ which will respect Eq.~(\ref{xi}) is given by:
\begin{eqnarray}
t(K)&=&\xi_L^{-1}(\xi_R(K))\ ,\ {\rm if}\ K>K^*,\\
&=&\xi_R^{-1}(\xi_L(K))\ ,\ {\rm if}\ K<K^*,\nonumber\\
&=&K^*\ ,\ {\rm if}\ K=K^*\nonumber\ .
\end{eqnarray}
The function $t(K)$ built in this way satisfies all the
requirements expected of it and spelled out above, and
in particular Eqs.~(\ref{consist}) and (\ref{xi}). 
Since both constructions are unique, they must be
equivalent to each other.

There is, of course, still a large number of further
applications and tests of the ideas presented here. 
To mention a few, one may for instance extend these considerations 
to systems which are self-dual, but whose dual lies on a different 
lattice (e.g., 2-d spins on a triangular lattice), or yet 
to systems which are simply {\it not} self-dual (e.g., 3-d Ising).
It is likely that new features of the RG would then come into play,
as is for instance the case with scheme dependence in the 
$\sigma$-model treated in \cite{H}. One may, furthermore, 
attempt to verify numerically the validity of any conjectures
resulting from duality considerations.

Finally, one further context in which duality consistency
relations may prove to be extremely useful is the quantum 
Hall effect (QHE). If we assume that the phase structure of the 
model is generated on the $(\sigma_{xx},\sigma_{xy})$
conductivity plane
by the modular group $SL(2,Z)$ or an appropriate subgroup thereof
(cf. \cite{L} for a detailed treatment of this RG picture), then
the same procedure as used above to derive Eq.~(\ref{consist})
yields immediately
some simple but interesting results. For instance, the 
procedure of applying $a(d/da)$ to a duality transformation
(in this case a modular transformation) and identifying 
the left hand side with the beta function evaluated at the dual point,
exactly as above, leads to the requirement that the beta function
transform as a modular function of weight $-2$. Such objects
are mathematically well-known and studied. Also, if we 
consider that all fixed points are generated by modular 
transformations of a finite number of elliptic points
in the fundamental domain,
the condition analogous to $t'(K^*)=-1$ ($t(K)$ is a modular
transformation, $K^*$ is an elliptic point) leads to the 
vanishing of the beta function at those points, as expected.
It is curious to note that in that case two consecutive 
duality transformations will in general {\em not} bring a
point back to itself (this is what enforced
$t'(K^*)=-1$ for us), and it is instead the existence of 
elliptic points in the fundamental domain which guarantee
the vanishing of the beta function at the fixed points.
Furthermore, while in the single-component case there is 
a single duality transformation relating two different
phases, in the QHE, there are an infinite number of duality
transformations, relating an infinite number of phases.
One may hope that such duality considerations, together with 
particular analyticity requirements, could eventually be 
stringent enough to determine the beta function completely.

We hope to report on some of these issues in the near future.

\vspace{0.5cm}
\noindent
{\sc Acknowledgement:} ~We thank M. Asorey and
C.B. Lang for discussions.



\begin{thebibliography}{X}
\bibitem{reviews}J. Kogut, Rev. Mod. Phys. {\bf 51} (1979) 659; \newline
R.D. Savit, Rev. Mod. Phys. {\bf 52} (1980) 453.
\bibitem{KW}H.A. Kramers and G.H. Wannier, Phys. Rev. {\bf 60} (1941) 252. 
\bibitem{B}T.H. Buscher, Phys. Lett. {\bf B194} (1987) 59; {\em ibid.}
{\bf B201} (1988) 466.
\bibitem{OQ}X. De La Ossa and F. Quevedo, Nucl. Phys. {\bf B403} (1993) 377.
\bibitem{H}P.E. Haagensen, Phys. Lett. {\bf B382} (1996) 356.
\bibitem{O}L. Onsager, Phys. Rev. {\bf 65} (1944) 117.
\bibitem{Ba}M.N. Barber, in {\em Phase Transitions and Critical Phenomena},
Vol. 8 (C. Domb and J.L. Lebowitz, Eds.), Academic Press (New York) 1983.
\bibitem{FB}M.E. Fisher and R.J. Burford, Phys. Rev. {\bf 156} (1967) 583.
\bibitem{Wu}F.Y. Wu, Rev. Mod. Phys. {\bf 54} (1982) 235;\newline 
H. Giacomini, Phys. Lett. {\bf A115} (1986) 13.
\bibitem{L}C.A.~L\"utken, Nucl.~Phys.~{\bf B396} (1993) 670, and references 
therein. Investigations along similar lines to the ones sketched here are
also being pursued by C.~Burgess and C.A.~L\"utken (C.~Burgess, private 
communication).

\end{thebibliography}
\end{document}